\documentclass[conference]{IEEEtran}
\IEEEoverridecommandlockouts
\usepackage{cite}
\usepackage{amsmath,amssymb,amsfonts}
\usepackage{graphicx}
\usepackage{textcomp}
\usepackage{xcolor}
\usepackage{booktabs}
\usepackage{tabularx}
\usepackage{array}
\usepackage[hidelinks,breaklinks=true]{hyperref}
\def\BibTeX{{\rm B\kern-.05em{\sc i\kern-.025em b}\kern-.08em
    T\kern-.1667em\lower.7ex\hbox{E}\kern-.125emX}}
\newcommand{\AEC}{\operatorname{AEC}}
\newcommand{\AEF}{\operatorname{AEF}}
\newcommand{\AUROC}{\operatorname{AUROC}}
\graphicspath{{figures/}}
\begin{document}

\title{Measuring Local Acoustic Contribution in Speech LLMs via Word-Span Counterfactual Interventions\\
}
      
\author{
\IEEEauthorblockN{Hardik B. Sailor$^{\dagger}$,
                  Qiongqiong Wang$^{\dagger}$,
                  Dheemant Rastogi$^{\star}$,
                  Nancy F. Chen$^{\dagger}$,
                  Ai Ti Aw$^{\dagger}$}
\IEEEauthorblockA{
$^{\dagger}$Institute of Advanced Intelligence and Computing, Agency for Science, Technology and Research, Singapore\\
$^{\star}$Nanyang Technological University, Singapore
}
}
\maketitle

\begin{abstract}
Correct transcription does not establish that a word depended on its local audio.
We introduce the Acoustic Evidence Contribution (AEC), a word-level counterfactual measure of acoustic grounding. It is obtained from the change in reference-token log likelihood when the word's aligned acoustic span is replaced. We combine decoding outcomes on unedited audio with thresholded AEC in an outcome–contribution matrix, where correct targets are categorized according to their local acoustic contribution as high or low. Depending on the system, 33--49\% of correct single-word targets fall below the twofold contribution threshold when averaged across corpora, despite no evident rank association with WER. Retention of reference words after span removal decreases monotonically across AEC quintiles, from 76--83\% in the lowest quintile to approximately 1\% in the highest. AEC therefore separates transcription correctness from dependence on the aligned local acoustic span while remaining an oracle-aligned diagnostic method.
\end{abstract}


\section{Introduction}

Modern automatic speech recognition (ASR) systems including multitask speech LLMs produce fluent
transcripts by combining acoustic observations with strong learned linguistic
regularities. Standard word or character error rates measure agreement with a
reference, but they do not reveal whether a correctly recognized word depended
on its local audio. Recent studies document fluent, acoustically unsupported ASR
outputs and show that aggregate error rates do not isolate this failure mode
\cite{koenecke2024careless,atwany2025lost,baranski2025nonspeech,koudounas2025hallucination,baranski2026halas}.
Word-level confidence estimation addresses the complementary problem of
predicting correctness \cite{qiu2021confidence}, but confidence alone does not
identify which acoustic evidence supported an output. This leaves a basic
measurement gap: correctness and local acoustic dependence are not the same
property.

Related work approaches acoustic grounding from several directions. ASR
hallucination detectors use common-output patterns, decoder states, or attention
features, while targeted head interventions reduce non-speech hallucinations
\cite{wang2025calmwhisper,jasinski2026internals,waldendorf2026attention}. Internal
language-model estimation suppresses acoustic components to expose linguistic
bias in autoregressive ASR and CTC systems \cite{meng2021ilme,das2023mask}, while
perturbation-based explanations search for audio frames sufficient for a
transcription \cite{wu2023explanations}. Audio-language evaluation further
separates text-prior solvability from audio reliance \cite{foo2026glitters};
multimodal trust estimation and contrastive decoding compare original audio with
equivalent, absent, blurred, or otherwise perturbed views
\cite{bhattacharya2025festa,hsu2025audioaware,li2026temporal,grace2026adaptive}.
Recently, Lebryk et al.\ \cite{lebryk2026benchmarkoptimization} quantify benchmark optimization at selected
positions. Lebryk et al. focus on specially constructed positions where the benchmark reference is underdetermined and test benchmark-specific recovery behavior. In contrast, we quantify the local acoustic contribution to arbitrary reference-aligned words, including normally recognized clean targets.
The resulting score measures counterfactual sensitivity to that local span.

We introduce \emph{Acoustic Evidence Contribution} (AEC), the mean decrease in
the teacher-forced log probability of a reference word's tokens when its aligned
acoustic span is replaced. Paired clean and edited passes keep the model, target,
tokenization, and prefix fixed, isolating the response to the local intervention.
Combining AEC with clean speech free-decoding outcomes yields the outcome--contribution
matrix, while a separate edited-audio free decode tests whether low- and
high-AEC targets behave differently after evidence removal. Code will be released upon acceptance.

\section{Method}

\subsection{Overview and notation}
 \vspace{-4mm}
\label{sec:notation}
\begin{figure}[h]
    \centering
    \includegraphics[width=1.0\linewidth, trim=0 10 0 20,clip]{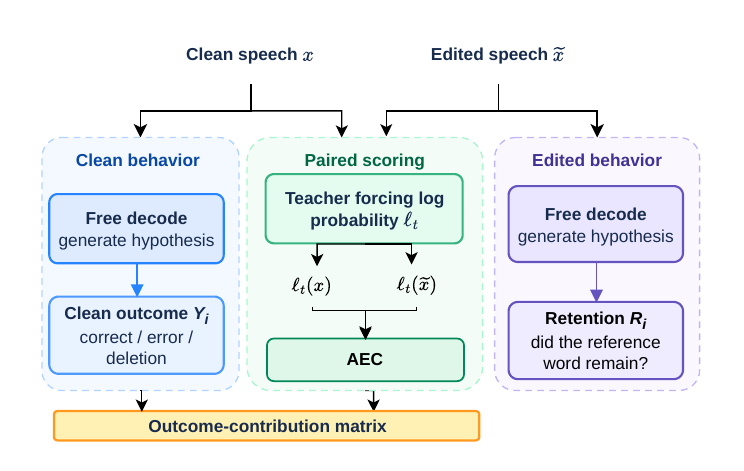}
    \vspace{-4mm}
    \caption{Three-part analysis for each aligned reference word or phrase span.}
    
    \label{fig:method_fig}
\end{figure}

For each target span, we examine three quantities:
(1) clean behavior: its correctness under clean speech decoding,
(2) paired scoring: the change in its teacher-forced reference likelihood when the aligned audio is edited,
and (3) edited behavior: its outcome under edited free decoding.
The second quantity measures local acoustic contribution, whereas the first and third
describe decoding behavior (Fig.~\ref{fig:method_fig}).

Let \(x\) denote a clean utterance with tokenized reference
\(\mathbf{b}=(b_1,\ldots,b_T)\).
Reference word \(w_j\) occupies token positions
\(\mathcal{T}_j\subseteq\{1,\ldots,T\}\), with
\(K_j = |\mathcal{T}_j|\). For a given evaluation set, let \(i\in\{1,\ldots,N\}\) index the \(N\) target interventions.
Intervention \(i\) covers a set of consecutive reference words
\(\mathcal{W}_i\) and their aligned acoustic interval \([s_i,e_i]\).
Replacing only this interval yields the edited utterance
\(\widetilde{x}_i\).
Thus, \(x\) and \(\widetilde{x}_i\) are complete utterances that differ only within
the target interval.

\subsection{Acoustic Evidence Contribution}
\label{sec:aec}

Under teacher forcing, the clean and edited passes use the same reference token,
reference prefix, and tokenization; only the audio input differs.
For reference token \(b_t\), we define the paired response to intervention \(i\) as log-likelihood difference,
\begin{equation}
    \Delta_{t,i}
    = \ell_t(x)-\ell_t(\widetilde{x}_i)=
    \log
    \frac{
        p_\theta\!\left(b_t \mid \mathbf{b}_{<t}, x\right)
    }{
        p_\theta\!\left(b_t \mid \mathbf{b}_{<t}, \widetilde{x}_i\right)
    }.
    \label{eq:token-response}
\end{equation}
This pairing avoids confounding the acoustic intervention with changes in the decoded
history.

The response of word \(w_j\) to intervention \(i\) is defined as
\begin{equation}
    \AEC_{j\leftarrow i}
    =
    (1/K_j)
    \sum\nolimits_{t\in\mathcal{T}_j}
    \Delta_{t,i},
    \label{eq:word-aec}
\end{equation}
and the Acoustic Evidence Contribution (AEC) of the target span is
\begin{equation}
    \AEC_i
    =
    (1/|\mathcal{W}_i|)
    \sum\nolimits_{j\in\mathcal{W}_i}
    \AEC_{j\leftarrow i}.
    \label{eq:target-aec}
\end{equation}

Averaging first over tokens normalizes each word for its number of subword tokens,
whereas averaging over target words gives equal weight to each constituent word of a
multi-word target.
For a single-word target,
\(|\mathcal{W}_i|=1\), and \(\AEC_i\) reduces to that word's response.

A positive \(\AEC_{j\leftarrow i}\) means that editing target span \(i\) decreases the
teacher-forced likelihood of reference word \(w_j\);
a value near zero indicates little sensitivity to the intervention;
and a negative value means that the edited audio assigns greater likelihood to the
reference word than the clean audio.
Hereafter, \(\AEC_i\) denotes the target-span quantity used in the
outcome--contribution analyses.
Off-target responses,
\(\AEC_{j\leftarrow i}\) for \(j\notin\mathcal{W}_i\),
are used to evaluate locality and to construct the untouched control distribution.

For interpretability, we define the Acoustic Evidence Factor (AEF) as $\AEF_{i}=\exp(\AEC_{i})$.
For a single-word target, $\AEF_{i}=\exp(\AEC_{i})$ is the geometric mean of the per-token clean-to-edited likelihood ratios. For a phrase target, it is the geometric mean of the constituent word-level evidence factors. Thus, $\AEF_{i}=2$ means that, on this normalized geometric-mean scale, the clean input assigns twice the target likelihood of the edited input. 
\subsection{Outcome--contribution matrix}
We use the interpretable threshold $\tau=\ln 2$ to define a high-contribution indicator for acoustic grounding:
\begin{equation}
g_{i}
 = \mathbf{1}\!\left[\AEC_{i}>\tau\right]
 = \mathbf{1}\!\left[\AEF_{i}>2\right].
\label{eq:evidence-label}
\end{equation}
Thus, $g_i=1$ indicates high local acoustic contribution, where the aligned span yields a target-level normalized likelihood factor greater than two; otherwise, $g_i=0$ indicates low local acoustic contribution.
Let $Y_i\in\{C,E,D\}$ denote whether the target associated with intervention $i$ is correct ($C$), substituted ($E$), or deleted ($D$) in the aligned clean
free decode. Throughout, we use \emph{correct target} to denote a target
recognized correctly in the clean free decode, i.e., $Y_i=C$. The matrix is reference-conditioned rather than hypothesis-conditioned; therefore insertion errors have no corresponding target span and are not represented. The resulting $3\times2$ cell counts define the outcome–contribution matrix:
\begin{equation}
n_{ys}
 = \sum_{i=1}^{N}
   \mathbf{1}\!\left[Y_i=y,\;g_{i}=s\right],
\quad
(y,s)\in\{C,E,D\}\times\{0,1\}.
\label{eq:cell-counts}
\end{equation}
The normalized cells $n_{ys}/N$ sum to one. Our primary summary
is the \emph{low-contribution correct share} (LCCS),
\begin{equation}
\mathrm{LCCS}
 := \Pr(g_{i}=0\mid Y_i=C)
 = n_{C0}/(n_{C1}+n_{C0}).
\label{eq:lecs}
\end{equation}
LCCS is the fraction of correct targets with low local acoustic contribution; higher values indicate weaker reliance on the aligned span. For a phrase target, the entire phrase is treated as one analysis unit. Its AEC is averaged over the constituent words and $Y_i=C$ only if every word is correctly recognized. We call this the span-macro analysis, in which each removed span whether a single word or a phrase contributes one observation. Our main analysis focuses on the $Y_i=C$ row; the remaining cells are reported in the supporting material \cite{sailor2026counterfactual}.
\subsection{Behavior after evidence removal}
The outcome--contribution matrix uses clean speech decoding and therefore does not
directly describe the model's behavior on the edited speech. We therefore freely decode the edited speech and align the resulting hypothesis to the original reference using minimum word-edit alignment. Let $R_{i}=1$ if the reference word/phrase is retained after the intervention otherwise $R_{i}=0$. Surplus hypothesis words do not alter $R_{i}$.

To test whether $\AEC$ has a graded behavioral relationship with
retention, we restrict the analysis to correct targets
(\(Y_i=C\)). Within each model--corpus cell, these targets are ranked by
AEC and partitioned into five equal-count bins. Let $Q_i\in\{1,\ldots,5\}$ denote the AEC quintile, from lowest $(Q_i=1)$ to highest $(Q_i=5)$. We measure
the retention rate in quintile \(q\) as $\Pr(R_i=1\mid Y_i=C,Q_i=q)$.

Rates are first computed within each model--corpus cell and then
macro-averaged over systems within each corpus. For a multiword target,
\(R_i=1\) only when every word in the target span remains correctly
recognized. This analysis tests whether retention decreases as local acoustic contribution increases. 
\section{Experimental setup}
\subsection{Datasets}
We constructed evaluation sets from LibriSpeech (test-other) \cite{panayotov2015librispeech}, AMI (IHM test set) \cite{carletta2006ami}, and TED-LIUM3 \cite{hernandez2018tedlium3} using Montreal Forced Aligner (MFA) \cite{mcauliffe17_interspeech}. Excluding spoken fillers, we sampled one target span per utterance ($\geq$6 eligible words, $\geq$3 non--target words remaining) from \{content, function, multiword (2--4 words), number, boundary, longest-duration\} with probabilities $(0.30,0.15,0.25,0.10,0.10,0.10)$. The evaluation contains 1,053 TED-LIUM, 5,181 AMI, and 2,693 LibriSpeech target spans. More details are provided in the supplementary material \cite{sailor2026counterfactual}. The aligned span was replaced by silence or duration-matched MUSAN noise/music \cite{snyder2015musan} with 20-ms cosine crossfades. We use silence as the primary intervention and noise/music as robustness checks. Unless stated otherwise, LCCS and retention analyses use silence. The same spans were reused across masking conditions using the same seed. 
\subsection{Models} We evaluate six models spanning 1.5B–24.3B parameters for this study. Four models are specialized ASR models: Whisper (1.5B)  \cite{radford2022whisper}, Qwen3-ASR (1.7B) \cite{shi2026qwen3asr}, Granite Speech 4.1 (2.3B) \cite{ibm2026granitespeech41}, and MERaLiON-3-ASR (3.3B) \cite{meralion2026meralion3asr}. We also selected two general purpose speech LLMs: Audio Flamingo Next (8.3B) \cite{ghosh2026audioflamingonext} and Voxtral Small (24.3B) \cite{liu2025voxtral}. We use greedy decoding for all models. 
For the speech LLMs, we use the prompt ``Transcribe the input speech, with no extra commentary.''
\subsection{Evaluation}
We report LCCS and retention rate for all six systems on the three corpora. We obtain 95\% confidence intervals from 1,000 utterance-level bootstrap resamples. Sensitivity of a score $S$ to the edited span is measured by $\mathrm{AUROC}_{\mathrm{span}}
= \Pr\!\left(S_i^{\mathrm{target}} > S_j^{\mathrm{untouched}}\right)$ where the target and untouched word are drawn from the same utterance. It is computed
separately for each corpus and intervention. Comparators include single-pass entropy, paired
clean--edited entropy change, and a silence-conditioned prior surprisal
defined as the reference-word negative log-likelihood conditioned on its
preceding tokens and silence-only audio.
\section{Results}
We organize the evaluation around three questions. RQ1 tests whether AEC
responds specifically to the edited span rather than to the utterance globally.
RQ2 estimates how often a correct target has low $\AEC$ at the
fixed threshold. RQ3 tests whether increasing AEC corresponds to progressively lower retention under edited-audio free decoding.
\subsection*{RQ1. Does AEC measure a localized response?}

\begin{figure}[h]
    \centering
    \includegraphics[width=1.0\linewidth]{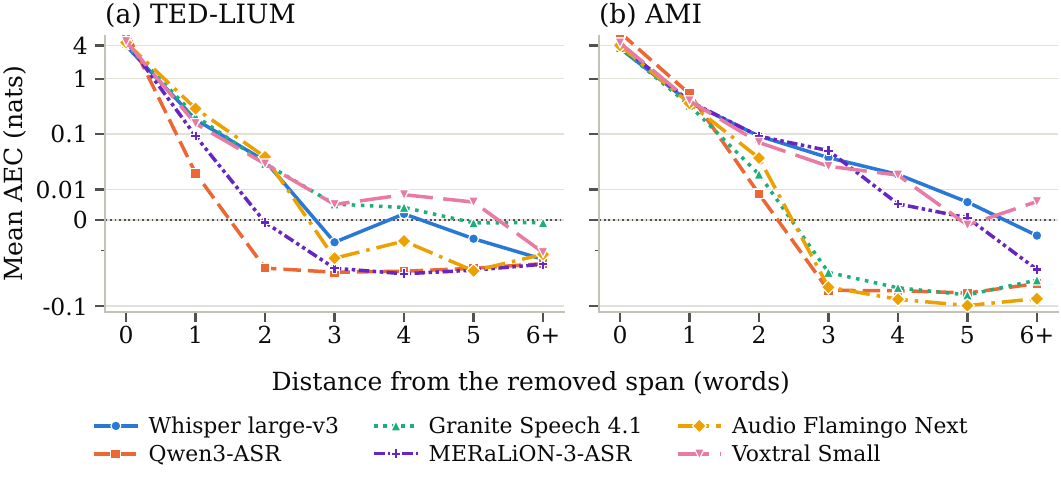}
    \vspace{-4mm}
    \caption{Spatial localization of AEC responses around the removed target span. LibriSpeech exhibits a similar localization pattern and is omitted for space.}
    \label{fig:aec-decay}
\end{figure}

Fig.~\ref{fig:aec-decay} plots $\AEC_{j\leftarrow i}$ as a function of word distance between \(j\) and the target intervention \(i\). It shows that the $\AEC$ response is concentrated
at the edited span and approaches zero for neighboring words.
This spatial decay indicates that $\AEC$ responds locally rather than
remaining elevated throughout the utterance. Although speech encoders can propagate a local input perturbation globally, the lower $\AEC$ on neighboring untouched words empirically supports that its output-level effect is strongly localized..

\begin{table}[h]
\centering
\caption{$\AUROC_{\mathrm{span}}$ avg. over the six systems.}
\label{tab:comparators}
\setlength{\tabcolsep}{2.4pt}
\renewcommand{\arraystretch}{1.08}
\begin{tabular}{@{}llccc@{}}
\toprule
Score & Edit info & Ref.-spec. &
Corpus (T/A/L) & Edit (S/N/M) \\
\midrule
Prior surprisal & None & \(\checkmark\) &
\(.58/.55/.54\) & \(.56/.56/.56\) \\
Entropy \(H\) & Edit only & \(\times\) &
\(.89/.82/.88\) & \(.85/.87/.87\) \\
\(\Delta H\) & Paired & \(\times\) &
\(.91/.85/\mathbf{.94}\) & \(.89/.90/.90\) \\
AEC & Paired & \(\checkmark\) &
\(\mathbf{.93}/\mathbf{.89}/.90\) & \(.90/.91/.91\) \\
\bottomrule
\end{tabular}
\vspace{1pt}
\parbox{\columnwidth}{\scriptsize
T/A/L: TED-LIUM/AMI/LibriSpeech; S/N/M: silence/noise/music.
``Ref.-spec.'' indicates whether the score explicitly uses the
probability assigned to the reference word. Corpus triplets avg. over the three interventions; intervention triplets avg. over the three corpora.}
\vspace{-1mm}
\end{table}
Table~\ref{tab:comparators} quantifies this localization through
$\AUROC_{\mathrm{span}}$. Prior surprisal, which does not observe the
local intervention, remains close to chance. Entropy from a single
edited pass performs better, while paired clean--edit scores provide
the strongest localization. AEC achieves the highest AUROC on
TED-LIUM and AMI and for each intervention family, whereas
$\Delta H$ is higher on LibriSpeech. The purpose of this comparison is not to identify the best localization score, but to verify that $\AEC$ responds robustly to the intervention. $\AEC$ and $\Delta H$ achieve comparable localization, but answer different questions: $\Delta H$  measures a change in distributional uncertainty, whereas $\AEC$ measures the directional change in probability assigned to the specified reference target. It therefore provides the reference-specific
and interpretable contribution axis required by the subsequent
outcome--contribution and retention analysis. Given the similar AEC localization across intervention families, subsequent analyses use silence removal unless stated otherwise.

\subsection*{RQ2. How often do correctly recognized words exhibit weak dependence on their aligned local acoustic span?}
Table \ref{tab:summary_low_AEC} reports LCCS under the primary silence intervention, with 95\% confidence intervals.
For single-word targets, 33-57\% of correct outputs have $\AEF \leq$ 2 on TED-LIUM, 29-43\% on AMI, and 33-46\% on LibriSpeech. The span-macro analysis shows the same general pattern, with lower absolute shares. These LCCS values show that correct recognition does not necessarily imply strong dependence on the aligned local audio. LCCS therefore identifies a substantial class of correct outputs that are only weakly sensitive to removal of their aligned span, without implying that they were generated from linguistic context alone. The clearer variation is across systems. Averaged over corpora, single-word LCCS ranges from 33.1\% for Audio Flamingo Next and 33.8\% for Qwen3 to 48.7\% for MERaLiON-3, a 15.6-point separation.

We also report the comparison of LCCS and WER in Table \ref{tab:wer_systems}.
This comparison is descriptive, not a quality ranking: low LCCS denotes stronger dependence on the specified local span, whereas WER measures recognition accuracy. Most usefully, it is not a restatement of accuracy: across the 18 model-corpus cells, there is no evident rank association between LCCS and WER (Spearman $\rho$ = $0.03$). Granite Speech 4.1 attains the lowest mean WER of the six systems ($5.2$\,\%) with one of the two highest LCCS values, and Audio Flamingo Next attains the lowest LCCS at a mid-field WER. LCCS therefore provides an acoustic contribution axis that WER does not capture.

\vspace{-2mm}
\begin{table}[h]
\centering
\setlength{\tabcolsep}{3.5pt} 
\renewcommand{\arraystretch}{1.15}
\caption{\% LCCS for single word and span-macro targets.}
\vspace{-2mm}
\label{tab:summary_low_AEC}
\begin{tabular}{l c c c}
\toprule
\textbf{System} & \textbf{TED-LIUM} & \textbf{AMI} & \textbf{LibriSpeech} \\
\midrule
\multicolumn{4}{l}{\textit{Single-word targets}} \\
Whisper-v3 (1.5B)      & $36.5 \pm 3.6$ & $37.5 \pm 1.7$ & $35.2 \pm 2.2$ \\
Qwen3-ASR (1.7B)       & $39.1 \pm 3.6$ & $28.9 \pm 1.6$ & $33.3 \pm 2.0$ \\
Granite Speech 4.1 (2.3B)     & $47.3 \pm 3.7$ & $43.0 \pm 1.7$ & $40.9 \pm 2.3$ \\
MERaLiON-3-ASR (3.3B)      & $57.1 \pm 3.5$ & $43.4 \pm 1.8$ & $45.7 \pm 2.2$ \\
Audio Flamingo Next (8.3B)  & $33.0 \pm 3.4$ & $33.9 \pm 1.6$ & $32.5 \pm 2.2$ \\
Voxtral Small (24.3B)  & $34.9 \pm 3.4$ & $36.7 \pm 1.6$ & $40.9 \pm 2.2$ \\
\textbf{Across models}& \textbf{33--57\%} & \textbf{29--43\%} & \textbf{33--46\%} \\
\midrule
\multicolumn{4}{l}{\textit{Span-macro targets}} \\
Whisper-v3 (1.5B)      & $28.8 \pm 3.0$ & $30.7 \pm 1.4$ & $28.8 \pm 1.8$ \\
Qwen3-ASR (1.7B)       & $29.8 \pm 2.9$ & $22.9 \pm 1.3$ & $26.6 \pm 1.8$ \\
Granite Speech 4.1 (2.3B)     & $37.6 \pm 3.0$ & $34.5 \pm 1.4$ & $33.2 \pm 1.8$ \\
MERaLiON-3-ASR (3.3B)      & $46.1 \pm 3.1$ & $35.3 \pm 1.5$ & $38.1 \pm 1.9$ \\
Audio Flamingo Next (8.3B)  & $25.3 \pm 2.7$ & $26.7 \pm 1.3$ & $26.1 \pm 1.8$ \\
Voxtral Small (24.3B)  & $27.7 \pm 2.9$ & $30.0 \pm 1.3$ & $33.9 \pm 2.0$ \\
\textbf{Across models}& \textbf{25--46\%} & \textbf{23--35\%} & \textbf{26--38\%} \\
\bottomrule
\end{tabular}
\end{table}

\vspace{-4mm}
\begin{table}[h]
\centering
\caption{
LCCS on single-word targets and unedited speech WER, per
corpus and averaged over the three.}
\vspace{-2mm}
\label{tab:wer_systems}
\setlength{\tabcolsep}{2.5pt}
\renewcommand{\arraystretch}{1.08}
\begin{tabular}{@{}lrcccc@{}}
\toprule
& & \multicolumn{2}{c}{LCCS (\%)} & \multicolumn{2}{c}{WER (\%)} \\
\cmidrule(lr){3-4}\cmidrule(l){5-6}
System & Par. & T/A/L & mean & T/A/L & mean \\
\midrule
Audio Flamingo Next & 8.3 & 33.0/33.9/32.5 & 33.1 & 3.9/11.8/3.8 & 6.5 \\
Qwen3-ASR & 1.7 & 39.1/28.9/33.3 & 33.8 & 4.4/13.3/4.0 & 7.2 \\
Whisper-v3 & 1.5 & 36.5/37.5/35.2 & 36.4 & 6.0/18.3/5.0 & 9.8 \\
Voxtral Small & 24.3 & 34.9/36.7/40.9 & 37.5 & 5.5/17.5/4.4 & 9.1 \\
Granite Speech 4.1 & 2.3 & 47.3/43.0/40.9 & 43.7 & 5.2/7.8/2.7 & 5.2 \\
MERaLiON-3-ASR & 3.3 & 57.1/43.4/45.7 & 48.7 & 4.0/15.8/4.6 & 8.1 \\
\midrule
\emph{Across models} & & \emph{41.3/37.2/38.1} & & \emph{4.8/14.1/4.1} & \\
\bottomrule
\end{tabular}
\end{table}
\vspace{-4mm}
\subsection*{RQ3. Does retention decrease as $\AEC$ increases?}
\begin{figure}[h]
    \centering
    \includegraphics[width=0.65\linewidth]{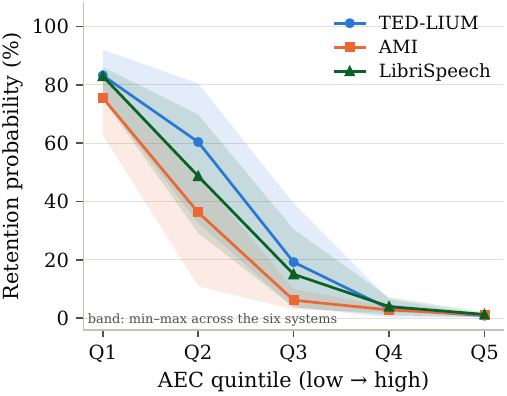}
    \vspace{-2mm}
    \caption{Retention after evidence removal as a graded function of AEC.}
    \label{fig:aec-quintile}
\end{figure}

Fig.~\ref{fig:aec-quintile} plots retention after evidence removal
against within-cell AEC quintiles. The bins are equal-count ranks within
each model--corpus cell, so the figure tests the ordering induced by $\AEC$
without selecting a threshold. 
It shows a strong graded relationship between
AEC and post-edit behavior. Retention decreases monotonically from the
lowest to the highest $\AEC$ quintile across 18 cells (6 models, 3 corpus).  The
six-system corpus means fall from 76--83\% in \(Q_1\) to
1\% in \(Q_5\). Across the 18 model--corpus cells, retention rates range from 2.5–17.3\% for high-AEC spans ($\AEF_i>2$) and from 69.3–96.1\% for low-AEC spans ($\AEF_i\leq2$). Thus, the $\AEC$ grouping predicts a distinct free-decoding behavioral response: a target with strong local contribution
usually disappears when its aligned evidence is removed.

\subsection*{Threshold calibration}
\begin{figure}[h]
     \centering
     \includegraphics[width=0.6\linewidth]{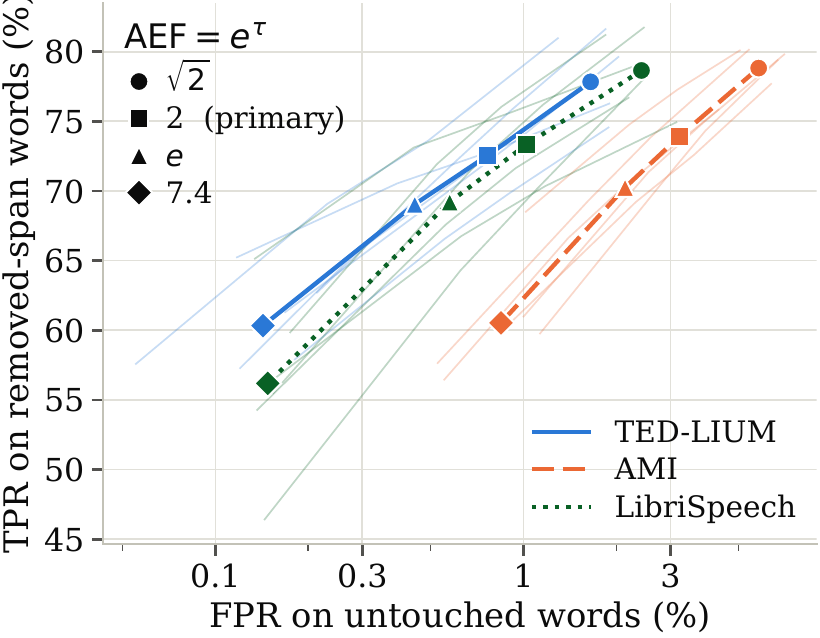}
     \vspace{-2mm}
     \caption{Calibration of the fixed AEF threshold. Thin lines represent
individual systems and thick lines represent corpus means.}
     \label{fig:threshold-calibration}
 \end{figure}

Fig.~\ref{fig:threshold-calibration} shows the False Positive Rate (FPR) on untouched words at distance $\geq2$ from the removed span against True Positive Rate (TPR) on removed span words as the threshold
varies over $\AEF\in\{\sqrt{2},2,e, e^{2}\approx7.4\}$.
At the prespecified $\AEF>2$ operating point, 69.8\,\% of target spans exceed the threshold while 1.33\,\% of untouched controls do, and no system--corpus cell exceeds 3.2\,\% false positives. The operating characteristics show that the interpretable $\AEF=2$ threshold provides a useful low-FPR operating point without being fitted to optimize this evaluation.
\vspace{-4mm}
\section{Conclusion}
We introduced AEC, a counterfactual word-level measure of local acoustic contribution, together with an outcome-contribution framework that separates transcription correctness from dependence on the aligned local acoustic span. AEC is strongly localized around the edited target and exhibits a graded behavioral relationship with post-intervention retention: correctly recognized targets become progressively less recoverable as local acoustic contribution increases. Across systems, the prevalence of low-contribution correct outputs varies substantially and is not explained by clean WER. These results show that conventional recognition accuracy does not capture how strongly a prediction depends on its local acoustic evidence. AEC therefore provides an interpretable oracle-aligned diagnostic for auditing local acoustic reliance in ASR and speech-language models.
\section*{Acknowledgment}
This research/project is supported by the National Research Foundation, Singapore under its National Large Language Models Funding Initiative. Any opinions, findings, conclusions, or recommendations expressed in this material are those of the author(s) and do not reflect the views of the National Research Foundation, Singapore.
\bibliographystyle{IEEEtran}

\bibliography{mybib}

\end{document}